\documentclass[fleqn,usenatbib,useAMS]{mnras}

\usepackage{graphicx}	% Including figure files
\usepackage{amsmath}	% Advanced maths commands
\usepackage{amssymb}	% Extra maths symbols
\usepackage{multicol}        % Multi-column entries in tables
\usepackage{bm}		% Bold maths symbols, including upright Greek
\usepackage{pdflscape}	% Landscape pages
\usepackage{multirow}

\usepackage[T1]{fontenc}
\usepackage{ae,aecompl}

\usepackage{newtxtext,newtxmath}
\title[SLSNe/LGRBs as birthplaces of FRBs]{Observing superluminous supernovae and long gamma ray bursts as potential birthplaces of repeating fast radio bursts}

\author[G. H. Hilmarsson et al.]{
G. H. Hilmarsson$^{1}$\thanks{E-mail: \href{mailto:henning@mpifr-bonn.mpg.de}{henning@mpifr-bonn.mpg.de}},
L. G. Spitler$^{1}$,
E. F. Keane$^{2}$,
T. M. Athanasiadis$^{1}$,
E. Barr$^{1}$,
\newauthor
M. Cruces$^{1}$,
X. Deng$^{3,1}$,
S. Heyminck$^{1}$,
R. Karuppusamy$^{1}$,
M. Kramer$^{1,4}$,
\newauthor
S. P. Sathyanarayanan$^{1}$,
V. Venkatraman Krishnan$^{1}$,
G. Wieching$^{1}$,
J. Wu$^{1}$
\newauthor
O. Wucknitz$^{1}$
\\
$^{1}$Max-Planck-Institut f{\"u}r Radioastronomie, Auf dem H{\"u}gel 69, D-53121 Bonn, Germany\\
$^{2}$SKA Organisation, Jodrell Bank Observatory, Lower Withington, Macclesfield, Cheshire, SK11 9FT, UK\\
$^{3}$CSIRO Astronomy and Space Science, Australia Telescope National Facility, PO box 76, Epping NSW 1710, Australia\\
$^{4}$Jodrell Bank Centre for Astrophysics, University of Manchester, Alan Turing Building, MP13 9PL, UK}
\date{Last updated 2015 May 22; in original form 2013 September 5}

\pubyear{2019}

\begin{document}
\label{firstpage}
\pagerange{\pageref{firstpage}--\pageref{lastpage}}
\maketitle

% Abstract of the paper
\begin{abstract}
Superluminous supernovae (SLSNe) and long gamma ray bursts (LGRBs) have been
proposed as progenitors of repeating Fast Radio Bursts (FRBs).
In this scenario, bursts originate from the interaction between a young
magnetar and its surrounding supernova remnant (SNR). 
Such a model could explain the repeating, apparently non-Poissonian
nature of FRB121102, which appears to display quiescent and active
phases.
This bursting behaviour is better explained with a Weibull distribution,
which includes parametrisation for clustering.
We observed 10 SLSNe/LGRBs for 63 hours, looking for repeating
FRBs with the Effelsberg-100~m radio telescope, but have not
detected any bursts.
We scale the burst rate of FRB121102 to an FRB121102-like source inhabiting each
of our observed targets, and compare this rate to our upper burst rate limit on a
source by source basis.
By adopting a fiducial beaming fraction of 0.6, we obtain 99.99\% and 83.4\%
probabilities that at least one, and at least half of our observed sources
are beamed towards us respectively.
One of our SLSN targets, PTF10hgi, is coincident with a persistent radio source,
making it a possible analogue to FRB121102.
We performed further observations on this source using the
Effelsberg-100~m and Parkes-64~m radio telescopes.
Assuming that PTF10hgi contains an FRB121102-like source, the probabilities of
not detecting any bursts from a Weibull distribution during our observations
are 14\% and 16\% for Effelsberg and Parkes respectively.
We conclude by showing that a survey of many short observations
increases burst detection probability for a source with Weibull
distributed bursting activity.

\end{abstract}

% Select between one and six entries from the list of approved keywords.
% Don't make up new ones.

%%%%%%%%%%%%%%%%%%%%%%%%%%%%%%%%%%%%%%%%%%%%%%%%%%

%%%%%%%%%%%%%%%%% BODY OF PAPER %%%%%%%%%%%%%%%%%%

% The MNRAS class isn't designed to include a table of contents, but for this document one is useful.
% I therefore have to do some kludging to make it work without masses of blank space.
%\begingroup
%\let\clearpage\relax
%\tableofcontents
%\endgroup
%\newpage

\section{Introduction}

Fast Radio Bursts (FRBs) are bright, highly dispersed, millisecond-duration radio transients
of unknown origin.
Since their inaugural detection \citep{2007Sci...318..777L}, 
close to 100 FRB discoveries have been published\footnote{\url{frbcat.org}} \citep{2016PASA...33...45P}. 
FRBs are believed to be extragalactic due to their high dispersion measures (DM),
which far exceed the expected Galactic DM contribution. This belief has strenthened
as FRBs have been increasingly
localized to host galaxies 
\citep{2017Natur.541...58C,Bannistereaaw5903,2019Natur.572..352R,2019Sci...366..231P,2020Natur.577..190M}.
While most FRBs detected so far have been single events,
FRB121102 was the first to be seen to repeat \citep{2016Natur.531..202S}, 
and recently nine repeating FRBs
have also been detected at CHIME \citep{2019Natur.566..235C,2019arXiv190803507T}. 
The repeating nature of some FRBs suggest that there are 
possibly two populations of FRBs, repeating and non-repeating.  

FRB121102 has been localized to a host galaxy \citep{2018Natur.553..182M}, and
its host identified 
as a low-metallicity dwarf galaxy at a redshift 
of $z=0.1927$ \citep{2017ApJ...834L...7T},
with a stellar mass of $\text{M}_* \sim 1.3{\times}10^8$ M$_\odot$ and a star formation rate of
0.23 M$_\odot$ per year \citep{2017ApJ...843L...8B}.
A compact persistent radio source with a projected size of <~0.7 pc 
was detected alongside FRB121102 \citep{2017ApJ...834L...8M}
and was determined to be co-located to within a projected 
distance of <~40 pc to the bursting source. 

Evidence for coincidence between FRB121102 and the persistent
radio source, along with the identification of the host galaxy, 
led to the suggestion of two types of progenitor models:
a magnetar wind nebula containing a young magnetar, 
embedded within a supernova remnant (SNR) \citep{2017ApJ...841...14M};
or a low luminosity active galactic nucleus (AGN) acting as the persistent
radio source, with the bursting activity either originating from the AGN
itself \citep{2016PhRvD..93b3001R}, or through interaction with a nearby neutron star (NS)
\citep[e.g.][]{2018ApJ...854L..21Z}.
In the case of FRB121102, the AGN model was initially thought unlikely, 
as dwarf galaxies rarely contain AGNs, along with the fact
that no evidence of an AGN in the optical spectrum was observed \citep{2017ApJ...834L...7T}.
However, a recent survey has shown that AGNs can be found offset
from the optical center of dwarf galaxies \citep{2020ApJ...888...36R}.
Additionally, the recently observed large and decreasing rotation measures (RMs) of FRB121102 ($\sim10^5$ rad/m$^2$) \citep{2018Natur.553..182M}, 
have drawn analogies between the system and
the Galactic center magnetar, J1745-2900 ($\sim-65000$ rad/m$^2$)
and Sagittarius A* system \citep{2018ApJ...852L..12D}. 

Supernovae occur from the collapse of massive stars into
black holes (BHs) or NSs. In some rare cases the remnant BH or NS powers
a relativistic jet into the circumstellar medium \citep{1993ApJ...405..273W},
and internal shocks within these jets can produce long gamma-ray bursts (LGRBs)
\citep{1994ApJ...430L..93R}.
Type-I superluminous supernovae (SLSNe) are a subclass of supernovae
which are hydrogen poor, orders of magnitude more luminous, 
have shorter decay times than the typical Type-I supernovae,
and have been postulated to be the precursor of LGRBs \citep{2019ARA&A..57..305G}.
The high luminosity is powered by a newly-born magnetar,
where the magnetar spin-dowon rate is tied to the short decay time \citep{2015Natur.523..189G}.
In addition to producing a fast-spinning NS with a strong magnetic field
that could produce more luminous radio bursts than Galactic NSs,
Type-I SLSNe and the resulting LGRBs also seem to occur more often
in low-mass, low-metallicity galaxies \citep{2006Natur.441..463F,2016ApJ...830...13P}.
Type-I SLSNe and LGRBs can therefore explain the repeating nature
of FRB121102 and its coincident persistent radio source. 
Note that throughout this paper, any mention of SLSNe is exclusively referring to Type-I SLSNe.

For FRB121102, the persistent radio source's luminosity is consistent with a model of radio emission from an SNR which is
powered by a young magnetar \citep{2017ApJ...841...14M}. A radio burst could therefore originate from the magnetosphere
of such a magnetar in a similar fashion to pulsar giant pulses \citep{2016MNRAS.457..232C}. 
Similarities in burst properties between FRB121102 and the Crab pulsar have been observed,
although whether giant pulses from the Crab can be scaled to the energies of FRB121102 is unclear \citep[see][and discussion therein]{2019ApJ...876L..23H}.
Alternatively, \citet{2019MNRAS.485.4091M} 
have modelled FRBs as synchrotron maser emissions from within an SNR. 
In that scenario, a central engine releases ultrarelativistic particles 
which collide with a mildly
relativistic magnetized ion-electron shell. The deceleration of the shell through forward shocks would
then produce FRBs through a synchrotron maser mechanism.
\citet{2019MNRAS.485.4091M} illustrate both production of FRBs within a large frequency range, 0.1--10 GHz, 
and the apparently dormant and clustering phases of bursts observed from FRB121102.

Shortly after the explosion, an SNR is optically thick at radio frequencies, 
so radio bursts from an embedded magnetar cannot be detected.
If the SNR is mainly ionized by the reverse shock, it can be probed at radio frequencies after a timescale of
centuries \citep{2016ApJ...824L..32P}.
However, assuming that along with the reverse shock of the supernova ejecta, the SNR is photoionized from within by the magnetar,
the SNR becomes optically thin at the frequency of the bursting emission
after $t\approx10 \;\nu_{\text{GHz}}^{-2/5}$ yrs \citep{2017ApJ...841...14M}.
At 1.4 and 6 GHz, $t$ is 8.7 and 4.9 yrs respectively, in the emitted frame. 

In this work we have identified and observed nine SLSNe and LGRBs as suitable sources for a targeted 
repeating FRB search at high frequencies (5.3--9.3 GHz) with the Effelsberg 100-m Radio Telescope.
The motivation for choosing this frequency range is that the SNR model allows for younger, and hence
more, sources to be observable; and that FRB121102 has been observed to emit at these frequencies \citep{2018ApJ...863....2G}.
We later added PTF10hgi to our 5.3--9.3~GHz survey and observed it during commissioning time
for the phased array feed (PAF) receiver at Effelsberg at 1.4 GHz 
and with the ultra wideband low (UWL) Parkes 64-m Radio Telescope receiver (0.7--4.0~GHz). 
This addition was made following the discovery of a radio source coincident with the SLSN PTF10hgi
at 6~GHz with the VLA \citep{2019ApJ...876L..10E}. This is the first detection of a persistent radio
source coincident with SLSNe/LGRBs, and it could be analogous to FRB121102's persistent radio source.
If an FRB were to be detected from PTF10hgi it would prove the theorised connection between FRBs and
SLSNe/LGRBs. 
Additionally, the age of PTF10hgi was roughly nine years at the time of observing, so its SNR should be only 
recently optically thin at 1.4~GHz. With our wide range of frequencies we could potentially observe
the optically thick-thin transition of the SNR.

Similar surveys have been performed recently: \citet{2019arXiv191002036L} observed 10 SLSN using the Karl G. Jansky Very Large Array (VLA)
for 8.5 hrs at 3 GHz, where they managed to detected the persistent radio source of PTF10hgi in their radio image searching.
\citet{2019MNRAS.489.3643M} observed five LGRBs and one short GRB for 20 hrs using the Robert C. Byrd Green Bank Telescope (GBT) at 820 MHz and 2 GHz, 
and the Arecibo Radio Telescope at 1.4 GHz. \citet{2019arXiv190911682M} observed six short GRBs, which originate from the 
merger of neutron stars and could leave behind a magnetar capable of producing repeating FRBs, for 20 hrs using the GBT at 2 GHz
and Arecibo at 1.4 GHz. No FRBs were detected in these surveys.

\section{Observations}
Our obsevations were carried out using The Effelsberg 100-m Radio Telescope
in Effelsberg,
Germany; and The Parkes 64-m Radio Telescope
in New South Wales, Australia.
The receivers used at Effelsberg were the S45mm single
pixel receiver, and the PAF; and at Parkes, the UWL receiver.
These will be described in their respective subsections below.  

The selection process for our targets was as follows. A list of SLSNe and LGRBs
was gathered from the Open Supernova Catalog\footnote{\url{https://sne.space}} \citep{Guillochon_2017} and the 
Swift GRB Catalog\footnote{\url{https://swift.gsfc.nasa.gov/archive/grb\_table.html}} with each source
being older than five years, 
and at a maximum redshift of 0.4. The age cut-off was conservatively set to five years 
to include only SNRs which are
optically thin in the observing band of the receiver.
The redshift
limit was set with respect to detections of 
FRB121102 at Effelsberg: 
By combining the radiometer equation \citep{1946RScI...17..268D} and
the brightness drop-off of the inverse square law,
it follows that a
detection with a signal to noise (S/N) of 40 at a redshift $z=0.2$ could 
be detected with a S/N of 10 at $z=0.4$ with the S45mm receiver.

The observed targets are
listed in Table \ref{tab:sources}, 
and the complete list of observations can be found in
Table \ref{tab:master}.
The range of our observations spans from June 2017 to September 2019. 
To reassure ourselves that the PAF system was working properly,
we observed a test pulsar, B1612+07, for
five minutes every hour during observations of PTF10hgi. 
To detect it we folded the test pulsar
data using \texttt{dspsr} from the pulsar analysis software library \texttt{psrchive}\footnote{\url{http://psrchive.sourceforge.net/}}. 
For 42.3 of our total 63 observing hours, we observed our original nine targets
with the S45mm receiver for 1--2~hrs each time with a $\sim$5 month cadence. 
We observed PTF10hgi for 5.3 hrs, split into two observations
of roughly 2.5 hrs each spaced a month apart with the S45mm receiver;
for 13 hrs with the PAF receiver for 1.5--4 hrs each day for four days;
and for 2.3 hours over three observations with the UWL spaced across six months.  
The strategy of multiple short observations was 
motivated by the apparent clustering of bursts from FRB121102 \citep{2018MNRAS.475.5109O}.

The total DMs of our potential radio sources can be broken down into individual contributions
by various components
\begin{equation}
\text{DM} = \text{DM}_\text{MW} + \text{DM}_\text{MWhalo} + \text{DM}_\text{IGM} + \text{DM}_\text{host} ,
\label{eq:dm}
\end{equation}
where $\text{DM}_\text{MW}$ and $\text{DM}_\text{MWhalo}$ are the DM contribution
of the Milky Way (MW) and its halo, respectively, $\text{DM}_\text{IGM}$ is the contribution of the intergalactic medium (IGM),
and $\text{DM}_\text{host}$ is the contribution of the host galaxy and the local environment of the source.
The $\text{DM}_\text{MW}$ varies between different lines of sight (LoS), but in general does not exceed 100 pc cm$^{-3}$ for
LoSs away from the Galactic plane, which is the case for most of our targets.
Using the Galactic electron density model YMW16 \citep{2017ApJ...835...29Y}, 
we obtain DM$_\text{MW}$ values between 22 and 143 pc cm$^{-3}$
for our targets.
We assume a DM$_\text{MWhalo}$ value of 50--80 pc cm$^{-3}$ \citep{2019MNRAS.485..648P}.
To estimate the DM$_\text{IGM}$ we use the relation $z\sim\text{DM}/855$ pc cm$^{-3}$ \citep{2018ApJ...867L..21Z} reulting in a
DM$_\text{IGM}$ range of 66-311 pc cm$^{-3}$.
The estimated DM$_\text{IGM}$ from recent FRB localisations are 
in agreement with this relation \citep{Bannistereaaw5903,2019Natur.572..352R}.
Note that LoS variations might vary from 100 to 250 pc cm$^{-3}$ for DM$_\text{IGM}$ for our
redshift range depending on models for halos' gas profile of ionized baryons \citep[Fig. 1, bottom panel]{2014ApJ...780L..33M}.
The $\text{DM}_\text{host}$ component can vary between FRB progenitor models,
types of host galaxies and local environments, orientation of the host galaxy, 
and the LoS to the source through its host \citep{2018arXiv180401548W}. 
The $\text{DM}_\text{host}$ estimate for FRB121102 is in the range of 55--225 pc cm$^{-3}$ \citep{2017ApJ...834L...7T}.
Using this range for our DM$_\text{host}$,
the estimated total DM of our targets falls in the range of 220--700 pc cm$^{-3}$.

\begin{center}
\begin{table*}
    \begin{center}
    \caption{Properties of the observed SLSNe/LGRBs.
            \textit{From left to right}:
            Source name, discovery date, right ascension (RA) and declination (DEC) in J2000 coordinates, sources redshift ($z$),
            the type of source, i.e. whether it's an LGRB or SLSN,
            and the average estimated total DM (pc~cm$^{-3}$).}
    \label{tab:sources}
    \begin{tabular}{c c c c c c c}
    Source name & Discovery date & RA   & DEC & $z$ & Type & DM [pc~cm${-3}$]\\ 
    \hline
    GRB050826 & 2005/08/26 & 05$^\text{h}$51$^\text{m}$02.6$^\text{s}$   & -02$^\circ$39$'$28.8$''$ & 0.297 & LGRB & 600\\ 
    GRB051109B & 2005/11/09 &23$^\text{h}$01$^\text{m}$52.6$^\text{s}$   & +38$^\circ$39$'$46.8$''$ & 0.080 & LGRB & 330\\ 
    GRB111225A & 2011/11/25 & 00$^\text{h}$52$^\text{m}$37.9$^\text{s}$   & +51$^\circ$34$'$22.8$''$ & 0.297 & LGRB & 590\\ 
    PTF09cnd & 2009/08/07 & 16$^\text{h}$12$^\text{m}$08.94$^\text{s}$   & +51$^\circ$29$'$16.1$''$ & 0.258 & SLSN & 450\\ 
    PTF10uhf & 2010/08/05 & 16$^\text{h}$52$^\text{m}$47$^\text{s}$   & +47$^\circ$36$'$21.76$''$ & 0.288 & SLSN & 480\\ 
    PTF10bjp & 2010/01/09 & 10$^\text{h}$06$^\text{m}$34$^\text{s}$   & +67$^\circ$59$'$19.0$''$ & 0.358 & SLSN & 550\\ 
    SN2010gx & 2010/03/13 & 11$^\text{h}$25$^\text{m}$46.71$^\text{s}$   & -08$^\circ$49$'$41.4$''$ & 0.230 & SLSN & 430\\ 
    PTF12dam & 2012/04/10 & 14$^\text{h}$24$^\text{m}$46.20$^\text{s}$   & +46$^\circ$13$'$48.3$''$ & 0.107 & SLSN & 320\\ 
    LSQ12dlf & 2012/07/10 & 01$^\text{h}$50$^\text{m}$29.8$^\text{s}$   & -21$^\circ$48$'$45$''$ & 0.250 & SLSN & 440\\ 
    PTF10hgi & 2010/05/15 & 16$^\text{h}$37$^\text{m}$47$^\text{s}$   & +06$^\circ$12$'$32.3$''$ & 0.099 & SLSN & 330\\ 
    \end{tabular}
    \end{center}
\end{table*}
\end{center}

\begin{center}
\begin{table*}
    \begin{center}
        \caption{List of observations in a chronological order for each source.
                \textit{From left to right}: 
                Source name, the dates and starting times for each observation (in UTC and MJD),
                observation duration, and the frequency range of the observation.}
        \label{tab:master}
        \begin{tabular}{c c c c c c}
Source name &   UT Date & UTC   & MJD   & Duration [min] & Frequency [GHz] \\ \hline
GRB050826         & 20170630  & 11:18:41  & 57934.47131   & 58 & 5.3--9.3\\
GRB050826       & 20171128  & 22:21:04  & 58085.93130   & 147 & 5.3--9.3\\
GRB050826        & 20180330  & 16:31:30  & 58207.68854   & 120 & 5.3--9.3\\
GRB050826        & 20181023  & 00:46:50  & 58414.03252   & 60 & 4.6--5.1\\
GRB050826    & 20181023  & 04:08:30  & 58414.17257   & 60 & 4.6--5.1\\ \hline
GRB051109B       & 20171128  & 16:03:34  & 58085.66194   & 120 & 5.3--9.3 \\
GRB051109B       & 20180330  & 09:23:50  & 58207.39155   & 90 & 5.3--9.3 \\
GRB051109B       & 20181022  & 17:00:20  & 58413.70856   & 60 & 4.6--5.1 \\ \hline
GRB111225A       & 20171128  & 18:12:54  & 58085.75896   & 120 & 5.3--9.3 \\
GRB111225A       & 20180330  & 11:06:51  & 58207.46309   & 46 & 5.3--9.3 \\
GRB111225A       & 20180330  & 18:35:50  & 58207.77488   & 51 & 5.3--9.3 \\
GRB111225A       & 20181022  & 18:12:00  & 58413.75833   & 60 & 5.3--9.3 \\ \hline
PTF09cnd         & 20170630  & 13:21:01  & 57934.55626   & 48 & 5.3--9.3 \\
PTF09cnd         & 20180330  & 05:11:30  & 58207.21632   & 120 & 5.3--9.3 \\
PTF09cnd         & 20181022  & 22:37:30  & 58413.94271   & 60 & 4.6--5.1 \\ \hline
PTF10uhf         & 20170630  & 15:17:01  & 57934.63682   & 55 & 5.3--9.3 \\
PTF10uhf         & 20180330  & 07:17:50  & 58207.30405   & 120 & 5.3--9.3 \\
PTF10uhf         & 20181022  & 21:36:40  & 58413.90046   & 60 & 4.6--5.1 \\ \hline
PTF10bjp         & 20170630  & 12:27:31  & 57934.51911   & 48 & 5.3--9.3 \\
PTF10bjp         & 20171129  & 01:26:14  & 58086.05988   & 120 & 5.3--9.3 \\
PTF10bjp         & 20180330  & 01:01:20  & 58207.04259   & 120 & 5.3--9.3 \\
PTF10bjp         & 20180330  & 14:08:10  & 58207.58900   & 18  & 5.3--9.3 \\
PTF10bjp         & 20181022  & 19:31:50  & 58413.81377   & 60 & 4.6--5.1 \\ \hline
SN2010gx         & 20171129  & 04:41:44  & 58086.19565   & 14 & 5.3--9.3 \\
SN2010gx          & 20180329  & 22:57:00  & 58206.95625   & 120 & 5.3--9.3 \\ \hline
PTF12dam         & 20170630  & 14:23:11  & 57934.59943   & 50 & 5.3--9.3 \\
PTF12dam         & 20171129  & 03:38:04  & 58086.15144   & 60 & 5.3--9.3 \\
PTF12dam         & 20180330  & 03:09:40  & 58207.13171   & 120 & 5.3--9.3 \\
PTF12dam         & 20181022  & 20:35:00  & 58413.85764   & 60 & 4.6--5.1 \\ \hline
LSQ12dlf         & 20171128  & 20:17:54  & 58085.84576   & 120 & 5.3--9.3 \\
LSQ12dlf          & 20180330  & 12:01:00  & 58207.50069   & 120 & 5.3--9.3 \\
LSQ12dlf         & 20181022  & 23:43:40  & 58413.98866   & 60 & 4.6--5.1 \\ \hline
PTF10hgi        & 20190205  & 18:48:31  & 58519.78369   & 42 & 0.7--4 \\
PTF10hgi        & 20190210  & 02:24:52  & 58524.10060   & 155 & 4--8 \\
PTF10hgi        & 20190220  & 20:15:55  & 58534.84439   & 44 & 0.7--4 \\
PTF10hgi        & 20190308  & 01:14:42  & 58550.05187   & 160 & 4--8 \\
PTF10hgi       & 20190323  & 23:54:29  & 58565.99618   & 216 & 1.222-1.452 \\
PTF10hgi       & 20190324  & 22:54:32  & 58566.95454  & 236 & 1.222-1.452\\ 
PTF10hgi       & 20190325  & 23:18:01  & 58567.97085  & 90 & 1.222-1.452\\ 
PTF10hgi       & 20190326  & 23:11:26  & 58568.96627  & 236 & 1.222-1.452\\
PTF10hgi        & 20190830  & 05:05:15  & 58725.21198   & 55 & 0.7--4 
        \end{tabular}
    \end{center}
\end{table*}
\end{center}

\subsection{S45mm receiver}
The S45mm receiver is located in the secondary focus of The Effelsberg Telescope,
and yields 4 GHz of bandwidth between either 4--8 GHz or 5.3--9.3 GHz.
The receiver has an SEFD of 18 Jy.
All the observations made using this receiver in this work are in the 5.3--9.3 GHz mode, except for
the observations of PTF10hgi, which were taken in the 4--8 GHz mode.
The data are recorded with full Stokes using two ROACH2 backends, each capturing
2 GHz of the band, with a 131 $\upmu$s sampling rate, and a 0.976562 MHz channel bandwidth
across 4096 channels. The resultant data are in a Distributed Aquisition and Data Analysis 
(DADA) format\footnote{\url{http://psrdada.sourceforge.net}}, from which Stokes \textit{I} is extracted.  

During the observaion on 22nd October 2018, a problem occurred with the S45mm receiver, resulting in
poor attenuation levels making the receiver 
temporarily
inoperable, and the use of a different
receiver was needed. The S60mm receiver on Effelsberg was used instead, 
with 500 MHz of bandwidth at 4.6--5.1 GHz, 82 $\upmu$s sampling rate,
512 channels with 0.976562 MHz bandwidth, and an SEFD of 18 Jy.
The data are recorded as sub-banded SIGPROC\footnote{\url{http://sigproc.sourceforge.net}} filterbanks, 
which are a stream of n-bit numbers corresponding to multiple polarization and/or frequency channels over time,
and are concatenated before processing.

\subsection{PAF receiver}
The Effelsberg PAF \citep{2018IAUS..337..330D}
is a dense array of antenna elements installed at the telescope's primary focus,
adapted from the models used by ASKAP \citep{2008RaSc...43.6S04H,2008ExA....22..151J}.
Its 188 elements form a checkerboard shape over a 1.2~m diameter circle and the output of these
elements are combined to form beams, controlled by varying the element weights. 

In its current state, the Effelsberg PAF can produce 22 beams, with 230 MHz of bandwidth centered at 1337 MHz,
and an SEFD of 34 Jy.
Currently the data are recorded and stored on disk as total intensity DADA files
with 512 channels of 0.449074 MHz bandwidth each, and a 216 $\upmu$s sampling time. 
The data can also be recorded as baseband data. This will be used in future surveys for real-time processing,
where we will use the raw voltage data captured from a ring buffer to create full Stokes files with
significantly higher frequency and time resolutions than our standard filterbanks.

\subsection{UWL receiver} 
The UWL receiver \citep{Dunning2015} is a wideband receiver at the Parkes telescope with an SEFD of 25 Jy.
It has a bandwidth of 3.3 GHz, ranging from 0.7 to 4 GHz. 
The data were recorded in two different modes with the MEDUSA backend: 
full Stokes, with a sampling time of 1024 $\upmu$s
and a channel bandwdith of 2 MHz across 1664 channels
for the first observation;
and Stokes \textit{I}, with a sampling time
of 256 $\upmu$s and 0.5 MHz channel bandwidth across 6656 channels for the latter two observations.
The 256 $\upmu$s data were downsampled by a factor of four for consistency and to reduce computation time
during analysis.
The data are in a Pulsar Flexible Image Transport System (PSRFITS) format \citep{2004PASA...21..302H}. %, which 

\subsection{Data processing}
All data products are initially converted to SIGPROC filterbank format before being processed.
For the S45mm, S60mm, and UWL data, 
the \texttt{PRESTO}\footnote{\url{github.com/scottransom/presto}} \citep{2011ascl.soft07017R} 
software package was used for single pulse searching. 
We used \texttt{PRESTO}'s \texttt{rfifind} to identify radio frequency interference (RFI)
in the data and create an RFI mask to apply to the data.
The data were dedispersed from 0--2000 pc cm$^{-3}$ in steps of 2 pc cm$^{-3}$ %,
for the S45mm and S60mm data, and in steps of 1 pc cm$^{-3}$ for the UWL data,
and subsequently searched
for single pulses using \texttt{PRESTO}'s \texttt{single$\_$pulse$\_$search.py} with a S/N threshold of 7.

\texttt{PRESTO} searches for single pulses by dedispersing the data and convolving the dedispersed
time series with boxcar filters of varying widths to optimise the S/N. 
\texttt{PRESTO} uses a pre-determined list of boxcar widths to use,
so by setting a maximum candidate width, \texttt{PRESTO} will search
using boxcars up to that width.
We search up to the nearest boxcar width of 20 ms, which is 19.6 ms.
We set this limit as FRBs tend not to have widths greater than a few ms
at our observed frequencies, 
and 20 ms is roughly the DM sweep in the S45mm band for the lower limit
of the estimated DMs of our targets.

We also compute the spectral modulation index of the candidates, which evaluates the fractional variation
of a candidate across its spectrum and distinguishes narrowband RFI
from broadband signals \citep{2012ApJ...748...73S}. 
The candidate's modulation index, $m_I$, is calculated as the normalized standard deviation of intensity across frequency,
and must be below the modulation index threshold,
\begin{equation}
\label{}
m_{I,\text{threshold}} = \frac{\sqrt{N_\nu}}{(\text{S}/\text{N})_\text{min}} ,
\end{equation}
where $N_\nu$ is the number of frequency channels, and $(\text{S}/\text{N})_\text{min}$ is the signal to noise
threshold applied to the data.
The candidates were then plotted and analysed by eye with a DM over time plot with marker sizes increasing with S/N.
Promising candidates were further inspected using \texttt{PRESTO}'s \texttt{waterfaller.py} plotting tool, which
shows the the candidate's dynamic spectrum and can be downsampled and subbanded at will.

For the PAF data, the GPU based single pulse search software 
\texttt{HEIMDALL}\footnote{\url{sourceforge.net/projects/heimdall-astro}} 
was used.
This was done to handle the vast amount of multibeam data taken, 
and to exploit \texttt{HEIMDALL}'s 
coincidencing capabilities.
\texttt{HEIMDALL}'s single pulse searching uses the same convolution method as \texttt{PRESTO}, 
but achieves much greater processing speeds by utilising GPUs rather than CPUs.
For the Effelsberg PAF, every frequency channel is calibrated independently,
and channels affected by RFI stronger than the calibration source have undefined
pointing positions, resulting in so-called badly beamformed channels.
A considerable portion of the channels in the PAF data needed to be 
zapped
during the processing due to both badly beamformed
channels, and channels persistently contaminated with RFI. 
These channels amounted to 89~MHz, or 39\% of
the PAF band, and were flagged to be ignored by \texttt{HEIMDALL}.
The data were dedispersed from 0--2000 pc cm$^{-3}$.
The DM steps in \texttt{HEIMDALL} are determined by the pulse broadening induced 
by the size of the DM step, so each DM trial is a function of the previous DM value and 
the data parameters \citep{2012PhDT.......306L}.
An initial detection threshold of $\text{S/N}=7$ was applied. 
\texttt{HEIMDALL} groups candidates which are close in DM and time,
and the group's candidate with the highest S/N is the candidate given by \texttt{HEIMDALL}.
This multi-beam data needed to be coincidenced in order to identify false candidates appearing
across many beams simultaneously, 
so the
single pulse candidates were ran through \texttt{HEIMDALL}'s \texttt{coincidencer}. The candidates were then
sifted further in order to reduce the large number of false positives
with low DMs and large widths: an increased S/N threshold of 8, 
a low DM threshold of 20 pc cm$^{-3}$, and a maximum candidate width of 28 ms were applied. 
In addition, candidates detected in mulitple beams go through further sifting. By taking the
beam with the strongest S/N as the reference point, the other beam detections need to occur 
within the adjacent beams for the candidate to pass the sifting.
The remaining candidates were then run through our own plotting tool\footnote{\url{github.com/ghenning/PAFcode}} which plots dedispersed time series,
dynamic spectrum, and a dedispersed dynamic spectrum. The dynamic spectra can also be downsampled and subbanded by factors
of our choosing. These plots were then inspected by eye.

We are aware of potential difficulties due to the DM sweep across the 4--8~GHz band. 
For a DM of 500 pc cm$^{-3}$ the sweep is 50~ms, so a narrowband signal might be
difficult to distinguish from zero-DM RFI. 
At the start of each observation we do however observe the pulsar B0355+54, which 
has a DM of 57 pc cm$^{-3}$, and are able to detect its single pulses.

\section{Results \& Analysis}
From the 63 hours of observational data, we have not detected any
single pulses 
from any of the sources observed
above our fluence limits of 0.04 $(w_\textrm{ms}/ \textrm{ ms})^{1/2}$ Jy ms
for the S45mm receiver, 0.53 $(w_\textrm{ms}/ \textrm{ ms})^{1/2}$ Jy ms
for the PAF receiver,
and 0.07 $(w_\textrm{ms}/ \textrm{ ms})^{1/2}$ Jy ms for the UWL receiver,
for burst widths of $w_\textrm{ms}$ ms. 

Assuming Poissonian statistics, we can estimate 
the upper-limit to the rate of bursts emitted above our detection threshold on a
source-by-source basis \citep[][Table 1]{1986ApJ...303..336G}. 
We also estimate the burst rate of an FRB121102-like source from each of the SLSNe/LGRBs observed.
The C-band results from the observed SLSNe/LGRBs and the PTF10hgi results with the PAF and UWL receivers 
are shown in their respective following subsections. 

To estimate the rate of an FRB121102-like source at different locations
we make use of a brightness distribution power-law,
\begin{equation}
\label{eq:lumdist}
R = R_0 \left( \frac{E}{E_0} \right)^\gamma,
\end{equation}
where $R$ and $E$ are the rate and energy, respectively, $R_0$ and $E_0$ are values
for a reference source, 
and $\gamma$ is the FRB brightness distribution power-law index. 
Here we use $\gamma=-0.91\pm0.17$ as estimated by \citet{2019MNRAS.486.5934J} independently of
instrumental sensitivity by combining the multi-telescope observing campaign of FRB121102 
\citep[1.4 and 3 GHz,][]{2017ApJ...850...76L}
and the GBT BL observations \citep[6 GHz,][]{2018ApJ...863....2G}.
An index of $\gamma = -1.8\pm0.3$ was obtained by \citet{2019ApJ...877L..19G} from 41 FRB121102 bursts at 1.4 GHz 
using Arecibo. 
These values of $\gamma$ are inconsistent with each other, potentially due to
Arecibo's survey probing unprecedentedly low burst energies of FRB121102, 
or its high sensitivity \citep{2019ApJ...877L..19G}.
We choose $\gamma=-0.91\pm0.17$ because the sensitivity of Effelsberg is
closer to GTB and VLA than Arecibo, and this value is partially derived from
detections in C-band.

The rate calculation also requires a relation between fluence and energy of a transient, 
considered specifically for the case of FRBs as \citet{2018MNRAS.480.4211M}
\begin{equation}
\label{eq:fluence}
F(\nu) = \frac{(1+z)^{2+\alpha}}{\Delta \nu_\text{FRB}} \frac{E}{4 \pi D^2_L},
\end{equation}
where $\Delta \nu_\text{FRB}$ is the intrinsic bandwidth of an FRB, 
$z$ is the source's redshift, $\alpha$ is the
spectral index, $E$ is the total energy of a burst, and $D_L$ is the luminosity distance
to the bursting source. 
The spectral index for FRBs is not well constrained, and given the absence of information
we assume a flat spectrum with $\alpha=0$ negating the need for a k-correction.
\citet{2019MNRAS.486.5934J} argues that Eq. \ref{eq:fluence} applies for 
bursts more broad-band than the observing bandwidth.
Bursts from FRB121102 have smaller fractional bandwidth
\citep[][Fig. 1]{2019ApJ...876L..23H}, 
so Eq. \ref{eq:fluence} can be written as the observed fluence averaged across the
observing band, $\Delta\nu$, \citep[][Eq. 8]{2019MNRAS.486.5934J}
\begin{equation}
\label{eq:fluence2}
F = \frac{(1+z)}{\Delta \nu} \frac{E}{4 \pi D^2_L}.
\end{equation}

We can then estimate the rate of bursts from 
FRB121102-like sources located at different luminosity distances/redshifts, for surveys with different
sensitivities by combining eqs. \ref{eq:lumdist} and \ref{eq:fluence2}:
\begin{equation}
\label{eq:rate}
R = R_0 \left(\frac{F}{F_0}\right)^\gamma \left(\frac{1+z_0}{1+z}\right)^{\gamma} \left(\frac{D_L}{D_{L,0}}\right)^{2\gamma} \left(\frac{\Delta \nu}{\Delta \nu_0} \right)^\gamma ,
\end{equation}
with the subscripts of 0 being the values for FRB121102, and $F_0$ being the fluence limit. 

\subsection{C-band observations of SLSNe/LGRBs}

\begin{figure}%[b]
    \centering
    \includegraphics[width=\columnwidth]{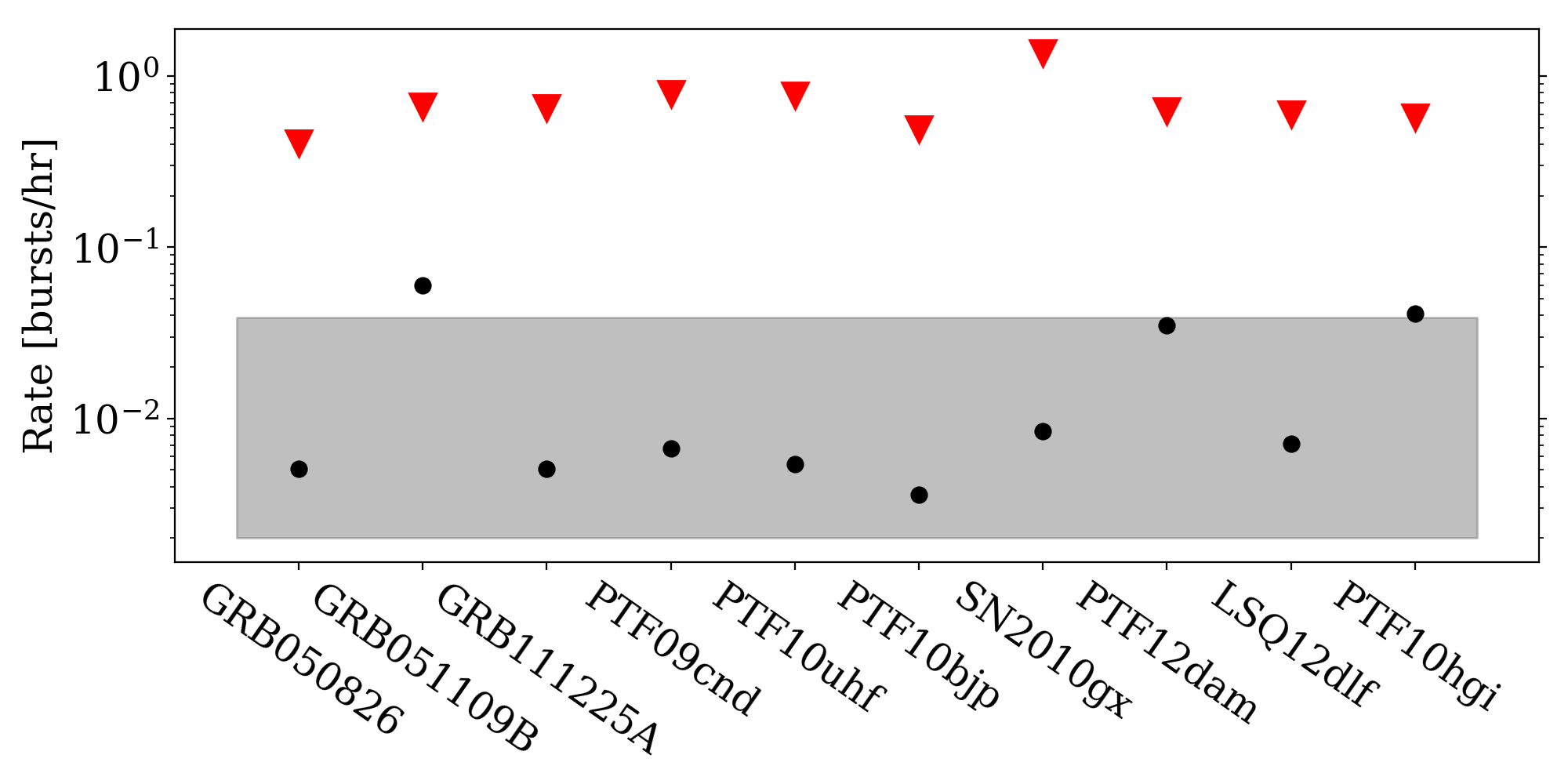}
    \caption{Scaled burst event rates of the observed sources at C-band.
            \textit{Gray bar}: 1$\upsigma$ burst event rate range of FRB121102 at C-band.
            \textit{Black dots}: Scaled burst event rate of an FRB121102-like source located at our SLSNe/LGRBs targets.
            \textit{Red arrows}: 95\% CL upper limit rates based on the non-detection of our observations.}
    \label{fig:results}
\end{figure}

The 95\% confidence level (CL) upper-limit to the burst rates we obtain from our observations, assuming Poissonian statistics,
are in the range of 0.4--1.4 bursts/hr and are shown in Table \ref{tab:results} and Fig. \ref{fig:results}.  

Observing campaigns of FRB121102 at high frequencies have reported various average burst rates.
\citet{2018ApJ...863....2G} reported 21 detections in a single 6 hr observation 
at the Green Bank Telescope (GBT) using the 4--8 GHz Breakthrough Listen (BL) Digital Backend. 
\citet{2018ApJ...863..150S} have three
detections in 22 hrs with the 4.6--5.1 GHz, S60mm receiver at Effelsberg in an observing campaign
spanning 4 months.

We also have obtained a rate of 0.012$^{+0.027}_{-0.010}$ bursts/hr 
(1$\sigma$ error)\footnote{All uncertainties in burst rates reported here are 1$\sigma$ errors.} 
from an ongoing campaign using
the 4--8 GHz, S45mm receiver at Effelsberg \citep{2020arXiv200912135H}. %(Hilmarsson et al., in prep).
In that campaign, which yields a single detection from 86 hrs of observations spanning two years, 
FRB121102 is observed for 2--3 hours at a time with a roughly two week cadence
(with gaps due to telescope/receiver maintenance). 
This rate is more robust than previously reported rates in the sense that it is a 
long-term average consisting of multiple observations, and does not depend on
a single bursting phase.
It is also obtained using the same observational setup as in this work. 

Using the burst rate of FRB121102 from \citet{2020arXiv200912135H}
%Hilmarsson et al. (in prep) 
of 0.012$^{+0.027}_{-0.010}$
bursts/hr, we estimate the burst rate of an FRB121102-like source located at each of the SLSNe/LGRBs observed.
Since we are working with the same observational
setup and identical bursts at different locations, 
we can simplify Eq. \ref{eq:rate} by setting $(F/F_0)^\gamma$ and 
$(\Delta \nu/\Delta \nu_{0})^\gamma$ to 1:
\begin{equation}
\label{eq:rate2}
R = R_0 \left(\frac{1+z_0}{1+z}\right)^{\gamma} \left(\frac{D_L}{D_{L,0}}\right)^{2\gamma}.
\end{equation}

The hypothetical rate of an FRB121102-like source located at our sources of interest
can be found in Table \ref{tab:results}, and is shown in Fig. \ref{fig:results}. There we have also estimated the number of
bursts we would have expected to see from an FRB121102-like source during our observations,
as well as how long we need to observe each source without a detection in order to constrain
our estimated rates, i.e. the observation time required for the upper limit to the rate to
reach the scaled rate.

The scaled rates from FRB121102 are 
influenced by the difference in luminosity 
distance between FRB121102 and the SLSNe/LGRBs,
yet they do fall within the 1$\upsigma$
range of FRB121102's rate at C-band.
This also implies that the time needed to constrain the scaled rates
reaches impractical observation times for most of the sources 
(upwards of 300 hours).
However, three of our sources, GRB051109B, PTF12dam, and PTF10hgi, have luminosity distances
less than FRB121102, and therefore have a higher scaled rate than FRB121102.
Since no bursts were detected in this work,
constraining the scaled rates of these three sources is quite a feasible task for further surveys. %,

\begin{table*}%[b]
    \begin{center}
    \caption{Results from the burst rate analysis of the observed sources.
            \textit{From left to right}:
            The observing band, source name,
            total observing time ($T_\text{obs.}$), 95\% confidence (CL) upper rate limit ($R_\text{UL}$),
            luminosity distance ($D_L$), the scaled burst rate from an FRB121102-like source ($R_\text{FRB121102}$),
            the expected number of bursts from our observations based on the scaled rates ($N_\text{exp.}$),
            and the observing time required to constrain the scaled rate ($T_\text{constr.}$;
            i.e. the time needed for the 95\% CL upper rate limit to reach the scaled rate).
            The luminosity distance and burst rate of FRB121102 are added for reference.
            \textit{Top}: S45mm receiver results. \textit{Center}: PAF receiver results.
            \textit{Bottom}: Parkes UWL results.}
    \label{tab:results}
    \begin{tabular}{c c c c c c c c}
    Obs. band [GHz] & Source name & $T_\text{obs}$ [hr] & $R_\text{UL}$ [hr$^{-1}$] & $D_L$ [Mpc] & $R_\text{FRB121102}$ [hr$^{-1}$]  & $N_\text{exp}$  & $T_\text{constr}$ [hr] \\ 
    \hline \\[-1.0em]
    \multirow{11}{*}{4--8} & FRB121102 & - & -  & 950     & 0.012$^{+0.027}_{-0.010}$ & -  & - \\
    &GRB050826 & 7.4 & 0.41  & 1550    & 0.01        & 0.04  & 582\\ 
    &GRB051109B & 4.5 & 0.67 & 370     & 0.06        & 0.27  & 50\\ 
    &GRB111225A & 4.6 & 0.65 & 1550    & 0.01        & 0.02  & 582\\
    &PTF09cnd & 3.8 & 0.79   & 1320    & 0.01        & 0.03  & 447\\
    &PTF10uhf & 3.9 & 0.77   & 1500    & 0.01        & 0.02  & 550\\
    &PTF10bjp & 6.1 & 0.49  & 1930    & 0.01        & 0.02  & 831\\
    &SN2010gx & 2.2 & 1.4   & 1150    & 0.01        & 0.02  & 358\\
    &PTF12dam & 4.8 & 0.62   & 500     & 0.03        & 0.17  & 85\\
    &LSQ12dlf & 5 & 0.60   & 1270    & 0.01        & 0.04  & 420\\
    &PTF10hgi & 5.3 & 0.57   & 460     & 0.04       & 0.21  & 73 \\
    \hline \\[-1.0em]
    \multirow{2}{*}{1.2--1.4} & FRB121102 & - &  -& 950 & 0.11$^{+0.04}_{-0.03}$  & - & - \\
    &PTF10hgi & 13.0 &  0.41$^{\text{a}}$& 460 & 0.4$^{+0.1}_{-0.1}$  & 5 & 14$^\text{a}$ \\
    \hline \\[-1.0em]
    \multirow{2}{*}{0.7--4.0} & FRB121102 & - &  -& 950 & 0.62$^{+0.33}_{-0.28}$  & - & - \\
    &PTF10hgi & 2.3 &  2.26$^{\text{a}}$& 460 & 2.2$^{+0.5}_{-0.5}$  & 5 & 3$^\text{a}$ \\
    \multicolumn{8}{l}{$^\text{a}$\footnotesize{99.5\% CL upper-limit.}}\\
    \end{tabular}
    \end{center}
\end{table*}

\subsection{PAF \& UWL observations of PTF10hgi}

We repeat the analysis from the previous section for the 13 hrs of PTF10hgi data taken with the PAF 
and the 2.3 hrs taken with the UWL.
In order to do so, we use the rate from a recent FRB121102 survey (28 bursts in 116 hrs) 
performed at L-band
using the P217mm 7-beam ($\text{SEFD}=17$ Jy) receiver at the Effeslberg telescope. 
The average burst rate is 0.24$^{+0.06}_{-0.05}$ bursts/hr above a fluence of
0.14 $(w_\textrm{ms}/ \textrm{ ms})^{1/2}$ Jy ms \citep{2020arXiv200803461C}. %(Cruces et al., in prep).

This rate must to be scaled to the PAF and UWL receivers, which we do using Eq. \ref{eq:rate}.
For the factor of $(F/F_0)^\gamma$ we use the radiometer equation \citep{1946RScI...17..268D}
\begin{equation}
\label{eq:radiometer}
F = \frac{\text{SEFD}\; \text{S/N}}{\sqrt{n_p \Delta \nu}}\sqrt{w} ,
\end{equation}
where SEFD is the system equivalent flux density, S/N is the signal to noise, 
$n_p$ is the number of polarizations, $\Delta \nu$ is the receiver bandwidth,
and $w$ is the burst width. The rate conversion from Eq. \ref{eq:rate} then becomes
\begin{equation}
\label{eq:rate3}
R = R_0 \left(\frac{\text{SEFD}}{\text{SEFD}_0}\right)^\gamma \left(\frac{\Delta\nu}{\Delta\nu_0}\right)^{\gamma/2} \left(\frac{1+z_0}{1+z}\right)^{\gamma} \left(\frac{D_L}{D_{L,0}}\right)^{2\gamma} ,
\end{equation}
where the last two bracketed terms are equal to 1 when converting rates for the same source.
The FRB121102 burst rate scaled to the PAF and UWL receivers is
$0.11^{+0.04}_{-0.03}$ bursts/hr above 
0.53 $(w_\textrm{ms}/ \textrm{ ms})^{1/2}$ Jy ms, and 
$0.62^{+0.33}_{-0.28}$ bursts/hr above 0.07 $(w_\textrm{ms}/ \textrm{ ms})^{1/2}$ Jy ms respectively.
Note that we are scaling burst energies to different observing bandwidths, so under the 
assumption that $R(E)$ does not depend on the central frequency of the observing bandwidth we add an
additional term of $\left(\Delta\nu/\Delta\nu_0\right)$ to Eq. \ref{eq:rate3}
when scaling from the P217mm receiver to the PAF and UWL receivers.

\begin{figure}%[b]
    \centering
    \includegraphics[width=\columnwidth]{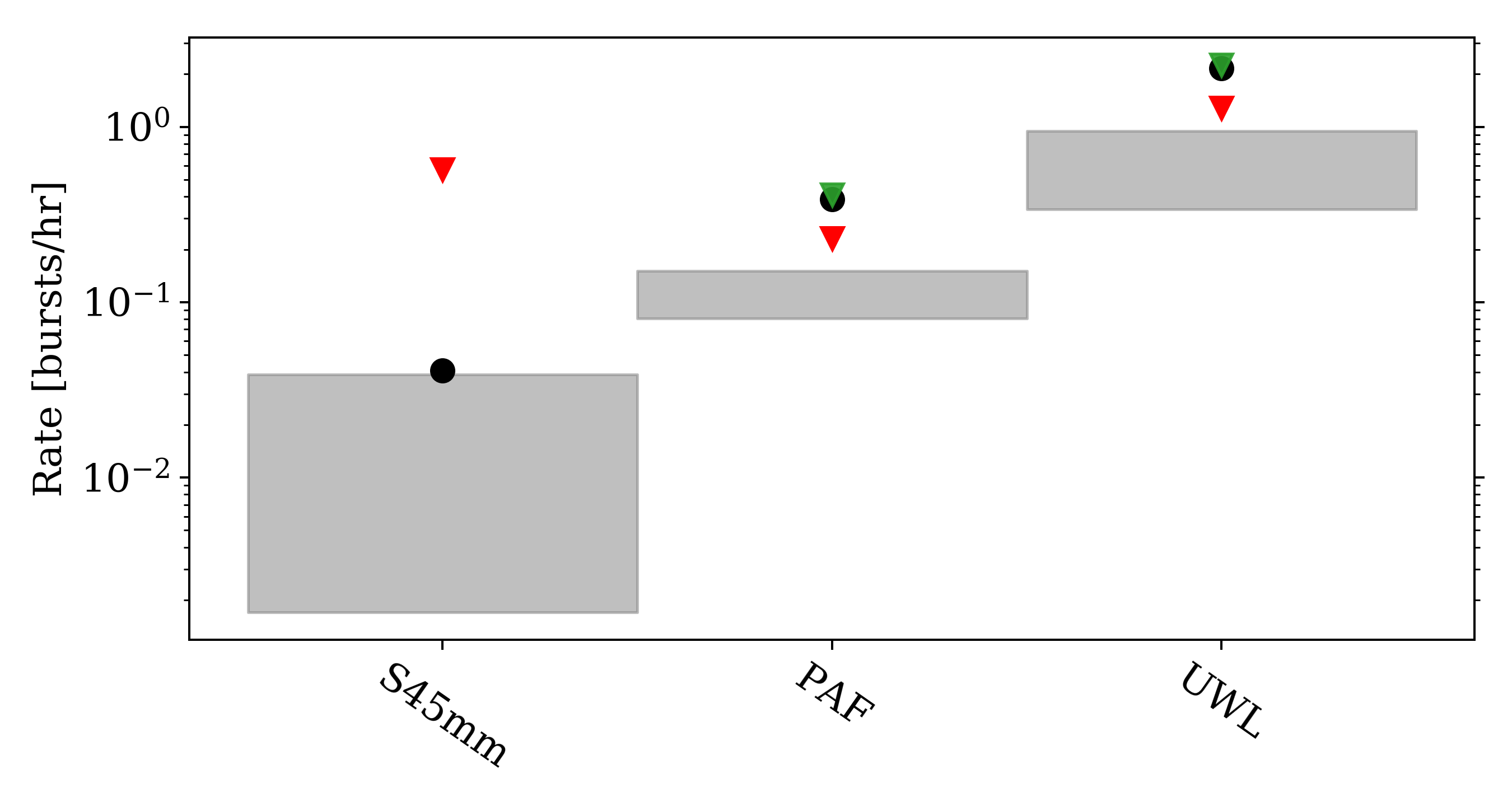}
    \caption{Scaled burst event rates of PTF10hgi for different receivers.
            \textit{Gray bars}: 1$\upsigma$ burst event rate range of FRB121102 at receiver.
            \textit{Black dots}: Scaled burst event rate of an FRB121102-like source
            located at PTF10hgi. 
            \textit{Red/Green arrows}: 95/99.5\% CL upper limits to the rates.}
    \label{fig:resultspaf}
\end{figure}

The resulting rate, obtained by using Eq. \ref{eq:rate3}, 
for an FRB121102-like source located at PTF10hgi
 is $0.4^{+0.1}_{-0.1}$ bursts/hr for the PAF receiver,
and $2.2^{+0.5}_{-0.5}$ bursts/hr for the UWL receiver. 
From the 13 hour observations with the PAF, 
we would have expected to detect 4--7 bursts on average by assuming this rate. 
We exclude this rate at the 99\% confidence level for such a source inhabiting PTF10hgi. 
Likewise, for the 2.3 hour observations with the UWL receiver we would have
expected 4--6 bursts on average, and exclude this rate at the 99\% confidence level.
The results are shown in the bottom section of Table \ref{tab:results} and in Fig. \ref{fig:resultspaf}.

The fact that we do not detect any bursts and rule out the rate of an FRB121102-like source
with a Poissonian distributed bursting activity
inhabiting PTF10hgi can be interpreted in various ways:
\textit{i)}~The most straightforward reason is that PTF10hgi simply does not contain
a repeating FRB source, or at the very least not a source as active as FRB121102, 
as FRB121102 might be an abnormally active bursting source \citep[e.g.][]{2018ApJ...854L..12P}.
\textit{ii)}~The assumption that FRBs are related to young magnetars within SNRs might not be correct.
\textit{iii)}~The FRB121102-like source may have been observed during a quiescent state, so no bursts were emitted during
our observations. If this were the case, it would directly imply that the bursting activity of the
source is non-Poissonian. 
\textit{iv)}~PTF10hgi's age was roughly nine years at the time of the observations,
so the SNR could be at the threshold of being optically thin at 1.4~GHz \citep{2017ApJ...841...14M}.
The SNR could simply still be opaque at 1.4~GHz, meaning that we cannot observe emitted bursts at that frequency,
given that the emission has to travel through the SNR.
\textit{v)}~The emission might be beamed and the bursts were not beamed towards us at the time of observing, 
so we were unable to detect bursts from the source. 

\subsection{Beaming fraction}
Emission mechanisms that generate luminous radio emissions are generally beamed, 
so a beaming fraction for our model should be taken into consideration.
The beaming fraction, $f$, is the fraction of the celestial sky covered
by the radio beam, and in the case of rotation it is how much is covered
during a single rotation.

The coherent emission process by a single unit
(particle or bunch of particles) has a beaming opening angle
of $1/\gamma_p$, where $\gamma_p$ is the Lorentz factor
of the unit.
\citet{2019ARA&A..57..417C} 
discuss three possibilities of FRB beaming geometries. 
First is a relativistic jet comprised of emission from multiple
incoherent units, 
and whose beaming is thus much greater than from the coherent
emission of a single unit.
Second is a relativistic jet rotating around an axis, where the
beam sweeps out an annulus shaped area during each revolution,
similar to pulsars.
Third is quasi-isotropic emission from a spherical shell.%}

Within the magnetar model framework, two distinct locations of
emission have been discussed in the literature:
a synchrotron maser mechanism from relativistic shocks
in the material surrounding the magnetar
\citep{2014MNRAS.442L...9L,2017ApJ...843L..26B},
and
pulsar-like emission in the magnetosphere
\citep{2017MNRAS.468.2726K,2018ApJ...868...31Y}.
\citet{2019MNRAS.485.4091M} 
model a synchrotron maser in a baryon-loaded shell that can produce bursts
over the full area of the SNR, relating to the aforementioned third beaming
gemoetry. 
The geometric probability of having a burst pointed towards an observer
is therefore 1.
Similarly, \citet{2019arXiv190807743B} proposes that FRBs are produced in an electron-positron plasma
in the helical-B winds of a rotating magnetar, where the geometric
probability is on the order of $\uppi$ steradians over the celestial sphere.
The lower limit to the beaming fraction of a burst is $1/\gamma_p^2$ 
for a single emitting unit. 
If multiple units are emitting, then the beaming fraction of a single FRB follows the
first beaming scenario previously described.
The probability of a burst being directed towards an observer depends
on the rate of burst generation and the beaming of each burst,
but the details are beyond the scope of this paper.%}

However, we can estimate beaming fractions relating to the second, 
pulsar-like beaming geometry using the Crab pulsar,
which has been used to model extragalactic FRBs \citep{2016MNRAS.457..232C}.
We estimate the beaming fraction of the pulsar-like emission
as a function of
the opening angle of the emission beam, $\uprho$, and
the angle between the rotation and magnetic axes, $\upalpha$
\citep[][Eq. 7]{1998MNRAS.298..625T}.
For the Crab, $\upalpha$ is estimated to be between 45 and 70 degrees
\citep{2013Sci...342..598L}, and $\uprho$ can be calculated from the 
pulsar's period \citep[][Eq. 7]{2001ApJ...553..341E}. 
The period of the Crab is 33.3~ms\footnote{\url{atnf.csiro.au/research/pulsar/psrcat/}} \citep{2005AJ....129.1993M}, 
resulting in $\uprho=20^\circ$. Thus the beaming
fraction of the Crab is between 0.5--0.7.

\begin{figure}%[b]
    \centering
    \includegraphics[width=\columnwidth]{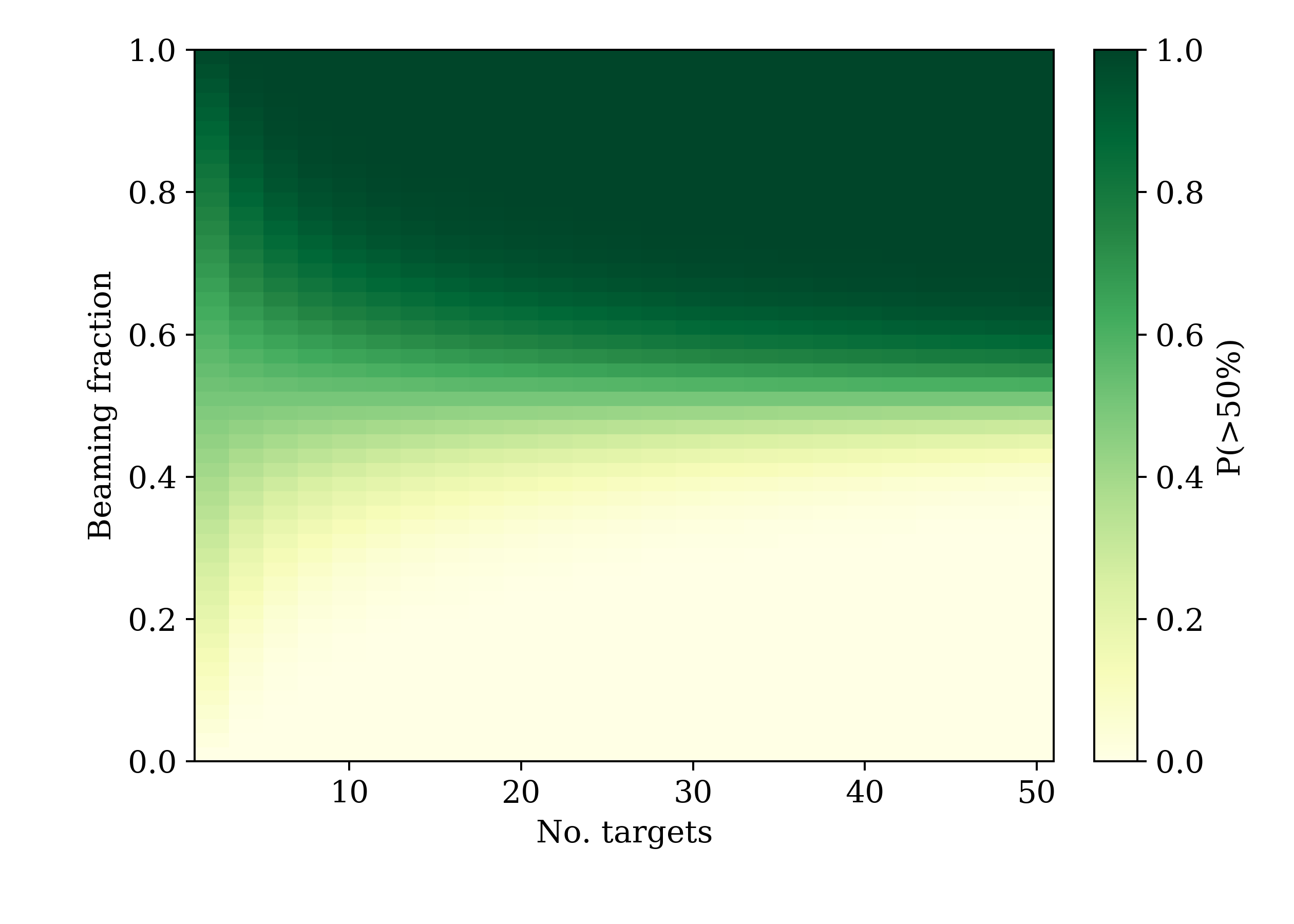}
    \caption{Probability that at least half of a list of observed 
            sources would be beamed towards us (P$>50\%$) as a function of 
            number of targets and beaming fraction.}
    \label{fig:beamfracprob}
\end{figure}

Assuming a fiducial beaming fraction value of 0.6,
there is a 99.99\% probability that at least one of our sources is beamed towards us
and an 83.4\% probability of at least half being beamed towards us.
We also plot the probability that half or more of targeted sources are 
beamed towards us, $P({>}50\%)$, as a function of various number of sources, $N$, and beaming fractions in Fig.
\ref{fig:beamfracprob}. From this figure we see that $P({>}50\%)$ consistently reaches above 70\% for $f>0.6$ and $N>10$.

\subsection{Poisson \& Weibull distributions}
Repeating bursts from FRB121102 have hitherto been treated as if
they follow the Poissonian process, which describes discrete, stochastically occurring events with
a known average time between them.
A Poisson distribution describes the probability to observe a number of
events following the Poisson process for an certain time period (e.g. an observation).

FRB121102 does not appear to follow this process. 
FRB121102 goes through phases
of quiescence and activity \citep{2016Natur.531..202S,2017ApJ...850...76L},
i.e. observed bursts appear clustered together.
A better way to describe bursts from FRB121102 might be with a Weibull distribution,
which has a more complex parametrization than a Poissonian distrubution. 
A Weibull distribution has a shape parameter, $k$, which describes
the degree of clustering; a rate parameter $r$;
and is written as \citet[][Eq. 2]{2018MNRAS.475.5109O}:
\begin{equation}
\label{eq:weibfull}
\mathcal{W}(\delta\mid k,r) = k\delta^{-1} \left[ \delta r \Gamma(1+1/k)\right]^\gamma e^{-\left[\delta r \Gamma(1+1/k)\right]^k} ,
\end{equation}
where $\delta$ are the intervals between subsequent bursts and $\Gamma$ is the gamma function.
For $k=1$, the Weibull distribution becomes
a Poissoinan distribution. If $k<1$, a clustering with small intervals
between bursts is favoured, so if a burst is detected, and observer is more likely
to detect subsequent bursts on a short timescale afterwards.
\citet{2018MNRAS.475.5109O} performed an analysis on L-band observations
of FRB121102 in order to estimate $k$ and $r$. They find that the 
posterior mean values of the shape parameter and rate are $k=0.34^{+0.06}_{-0.05}$
and $r=0.24^{+0.13}_{-0.08}$ bursts/hr, respectively. 
They also find that the Poissonian case of $k=1$ is strongly disfavored.
%Cruces et al. (in prep) 
\citet{2020arXiv200803461C} performed the same analysis on their aforementioned survey,
and obtain a shape factor of $k=0.39^{+0.05}_{-0.03}$ and a rate of $r=0.27^{+0.09}_{-0.08}$ bursts/hr.
These Weibull analysis rates are consistent with the Poissonian rate of $0.24^{+0.06}_{-0.05}$ bursts/hr 
above 0.14 $(w_\textrm{ms}/ \textrm{ ms})^{1/2}$ Jy ms
from \citet{2020arXiv200803461C}. %Cruces et al. (in prep).

We can estimate the probability of not detecting a burst from a source with
Weibull-distributed bursting activity for an observation of duration $\Delta_\text{obs}$ as \citet[][Eq. 18]{2018MNRAS.475.5109O}:
\begin{equation}
\label{eq:weibull}
P(N=0\mid k,r) = \frac{\Gamma(1/k) \Gamma_i\left(1/k,(\Delta_\text{obs} r \Gamma(1+1/k))^k\right)}{k \Gamma(1+1/k)} ,
\end{equation}
where $r$ is the bursting rate, 
$\Gamma$ is the gamma function, and $\Gamma_i$ is the incomplete gamma function.
The likelihood for multiple observations can be obtained by multiplying the probabilities
of each individual observation, given that the cadence of the observations is greater
than the spacing between bursts.

We can estimate this probability for our 13 hr PAF observations at L-band of PTF10hgi
(consisting of four separate observations of 1.5--4 hrs, see Table \ref{tab:master}).
We use the values from \citet{2020arXiv200803461C}
%Cruces et al. (in prep) 
of $k=0.39$ and $r=0.27$ bursts/hr.
First we need to scale this rate from FRB121102 to PTF10hgi 
and from the P217mm receiver to the PAF receiver 
using Eq. \ref{eq:rate3}, resulting
in a rate of $r=0.43$ bursts/hr. 
The resulting probability of not detecting a burst from these observations, assuming that
PTF10hgi contains an FRB121102-like source, is 14\%.

We repeat this analysis for the UWL observations, 
which were three observations of 42, 44, and 55 minutes,
using the same shape factor and rate.
The scaled rate of PTF10hgi from the P217mm to the 
UWL receiver is 2.42 bursts/hr, and we obtain a 16\%
probability of not detecting a burst from these observations.

\begin{figure}%[b]
    \centering
    \includegraphics[width=\columnwidth]{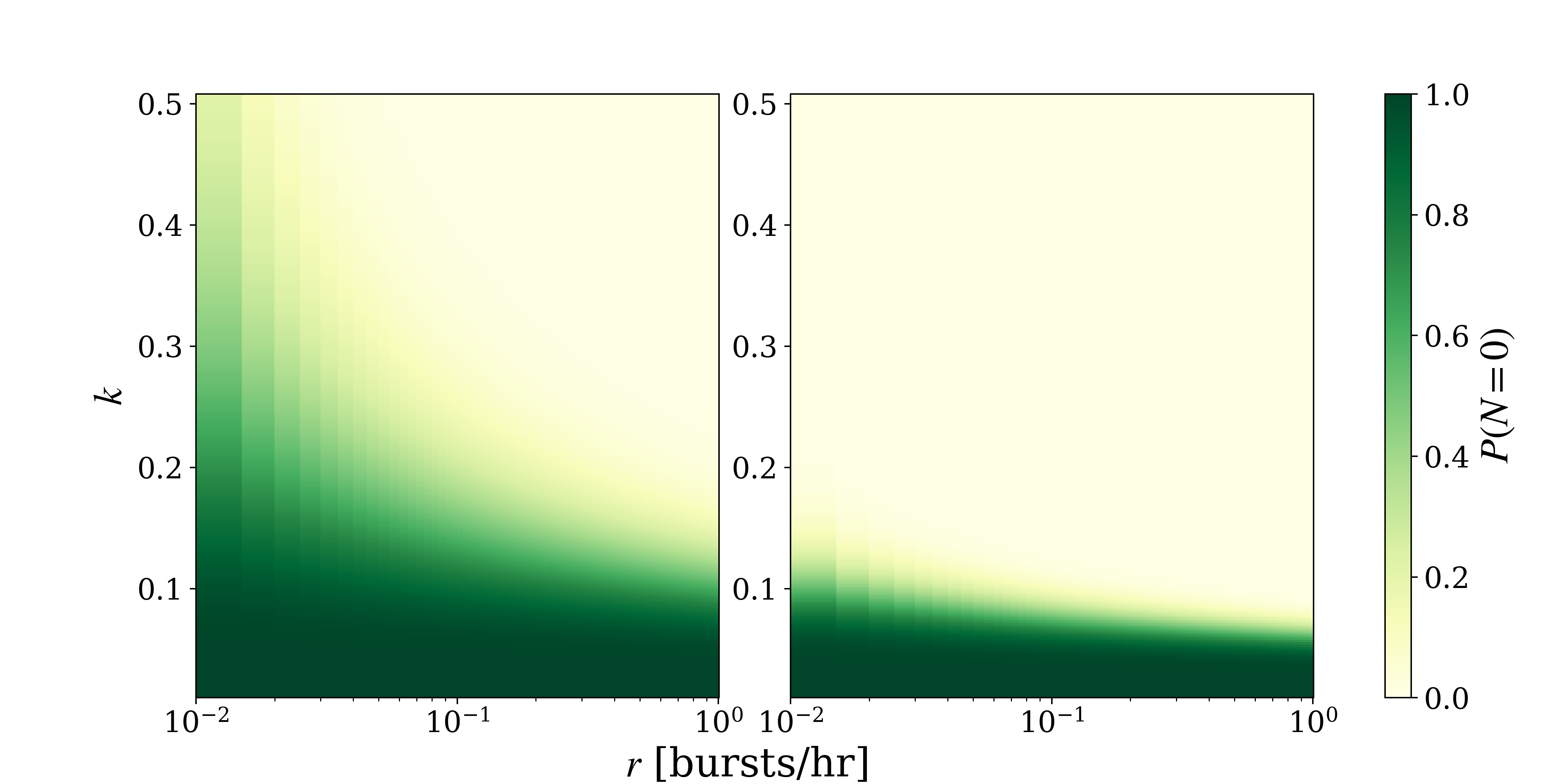}
    \caption{Probability of detecting zero bursts, $P(N=0)$, for a Weibull distributed
            bursting activity of a source as a function of shape parameter, $k$, and rate $r$.
            \textit{Left}: Probabilities from our two (5.3 hr in total) observations 
            of PTF10hgi at C-band. \textit{Right}: Probabilities from a hypothetical 72 hour
            observing campaign ($24\times3$ hrs).}
    \label{fig:Pk}
\end{figure}

To perform the same calculations for the S45mm receiver observations, a burst event
analysis for C-band observations of FRB121102 is needed in order to estimate the
shape parameter $k$ and rate $r$. This analysis is beyond the scope of this work, 
however we plot the probability of detecting zero bursts
as a function of the shape parameter and rate. 
We illustrate two cases: the observations of PTF10hgi with the S45mm receiver 
presented here, and a hypothetical survey
of 24 3-hr sessions (i.e. the time required to constrain the upper rate limit,
see Table \ref{tab:results}), shown in Fig. \ref{fig:Pk}.
There we see that the probability of not detecting
any bursts rapidly decreases with increasing $k$,
and that we can already exclude the L-band parameters
from our observations. 
The lack of bursts detected from 
FRB121102 at C-band, compared to L-band detections, might lead one to believe that
$k$ and $r$ are frequency dependent, with both being
lower at higher frequencies.
If we were to continue obsering PTF10hgi at C-band in the same fashion until we have
reached the time to constrain the Poisson rate,  
we could also place constraints on $k$ and $r$. 
This hypothetical survey of 72 hrs shows
that there is 0\% chance of not detecting a burst 
for $k\geq0.2$ and $r\geq0.03$ burst/hr, and would
constrain the upper-limit of $k$ to ${\sim}0.2$ if
no burst was detected.

\section{Discussion \& Conclusions}
In this work we investigate the possibility of SLSNe/LGRBs hosting
FRB121102-like progenitors.
We have observed 10 targets for 63 hours 
using the S45mm (5.3--9.3 GHz) and PAF (1.2--1.5 GHz) receivers at Effelsberg
and the UWL receiver (0.7--4 GHz) at Parkes,
but have found no bursts. 

By assuming an FRB121102-like source is located at our observed targets,
we have estimated their scaled burst rates with respect to luminosity distance,
redshift, and telescope sensitivity. We have also calculated the upper limit rate
for each source, based on our non-detections. The rate upper limits do
not constrain any of the scaled rates at C-band, but the scaled rates 
for three of our sources, GRB051109B, PTF12dam, and PTF10hgi, can be 
constrained with a reasonable amount of observing time.

PTF10hgi is a source of particular interest, as a persistent radio source
which is coincident with the SLSN was recently detected. This system could
be analogous to FRB121102, and detecting an FRB originating from it could be instrumental
in deciphering the enigmatic nature of these bursts. We have therefore spent
5.3 hrs observing PTF10hgi at 6 GHz with the S45mm receiver, 
13 hrs at 1.4 GHz during the commissioning of the
PAF receiver at Effelsberg, 
and 2.3 hrs at 2.4 GHz with the UWL receiver at Parkes. 
We did not detect any bursts from those observations,
and rule out at the 99\% CL the scaled PAF and UWL rates at L-band of an 
FRB121102-like source inhabiting PTF10hgi.
There are several possibilities for why we have not detected any bursts: 
\textit{i)}~PTF10hgi does not contain an FRB121102-like source,
\textit{ii)}~FRBs might not be related to young magnetars within SNRs,
\textit{iii)}~the source was observed during a quiescent state,
\textit{iv)}~PTF10hgi's SNR might still be opaque at L-band,
\textit{v)}~or bursts from the source are simply not beamed towards us,

When we adopt a beaming fraction of 0.6 for our sources we show
there is 99.99\% chance that at least one of our hypothetical targets would be
beamed towards us, and an 83.4\% probability that at least five
of them are beamed towards us. From Fig. \ref{fig:beamfracprob} we note that
for beaming fractions larger than 0.6,
at least half of the sources will consistently have a high probability of
being beamed towards us.

The clustering of bursts from FRB121102 could be better explained with a Weibull
rather than a Poissonian distribution \citep{2018MNRAS.475.5109O}. 
Using a shape factor of $k=0.39$ and a scaled rate of 0.43 bursts/hr for a Weibull distribution
\citep{2020arXiv200803461C}
%(Cruces et al., in prep)
we estimate
a 14\% probability of not detecting a burst from our PAF receiver observations of
PTF10hgi, 
assuming it contains an FRB121102-like source. 
By using the same shape factor and a rate of 2.42 bursts/hr for our UWL observations
we estimate a 16\% probability of not detecting a burst.
We do not have an
estimate of the shape factor at C-band, however we plot the probability of not
detecting a burst as a function of $k$ and $r$ for our S45mm receiver observations of 
PTF10hgi in Fig. \ref{fig:Pk}, and show that the L-band rate and shape factor
are already excluded.

\begin{figure*}%[b]
    \centering
    \includegraphics[width=\textwidth]{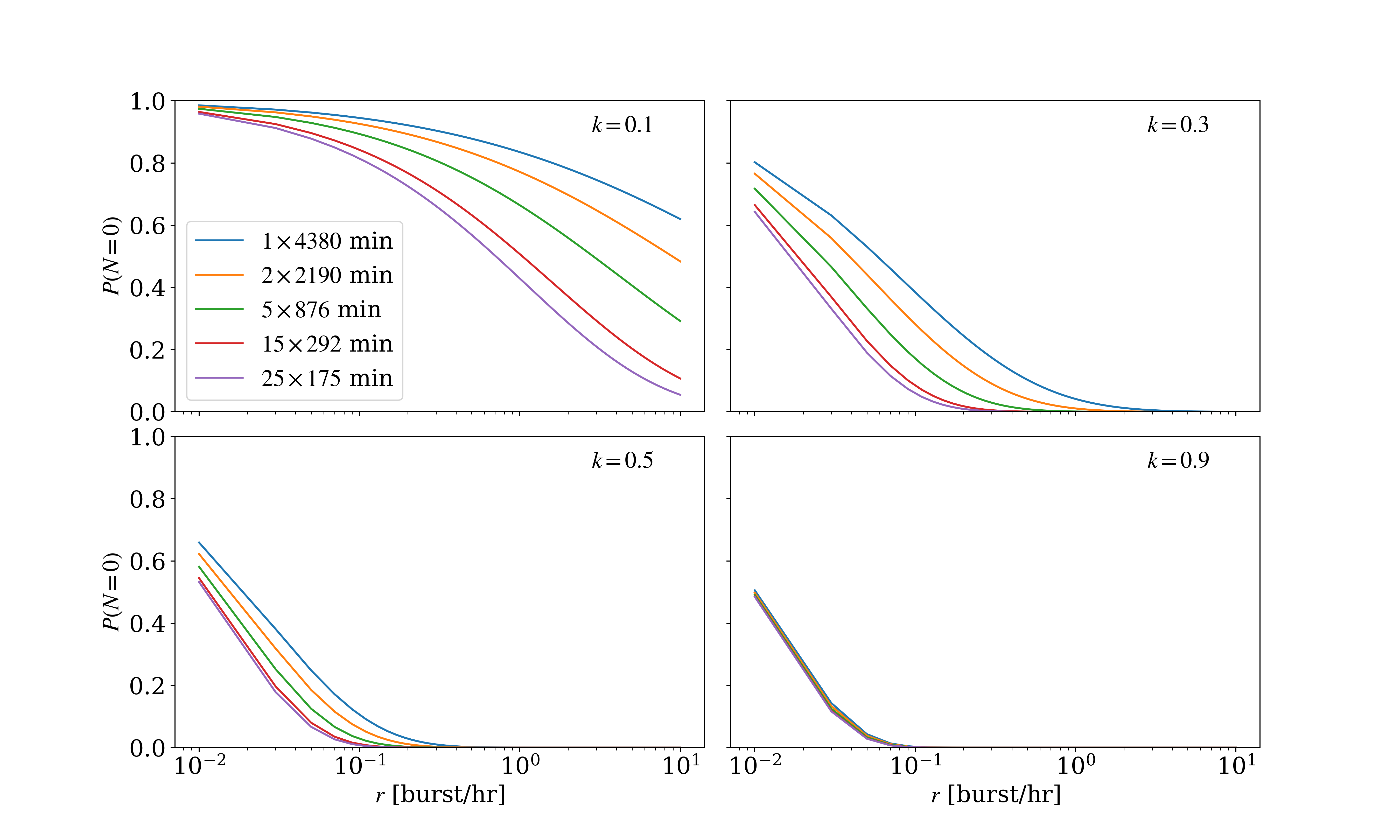}
    \caption{Probability of detecting zero bursts as a function of burst rate 
            for hypothetical 73 hr surveys spanning 
            different numbers of observations (colours). 
            Different shape factors, $k$, are shown in each panel.
            As expected a single long observation has the highest probability
            of detecting zero bursts from a Weibull distribution, with the
            probability decreasing with number of observations.}
    \label{fig:P0}
\end{figure*}

We have several recommendations for future surveys which may follow up this work.
By assuming that SLSNe/LGRBs contain FRB121102-like sources, we must expect
that they also have clustering of emission, along with periods of dormancy.
We should also assume that the bursts are beamed to some degree.
Therefore we suggest that observing multiple sources for short periods of time
on a regular basis would be ideal. 
The advantage of observing a source which has clustered burst phases across multiple
short observations rather than a few (or one) long observations is shown in Fig. \ref{fig:P0}.
There we plot the probability of detecting zero bursts for a survey totaling 73 hrs across
different number of observations as a function of burst rate for a source with different shape factors.  
As we move further away from the Poissonian case of $k=1$, it becomes increasingly important
to split a survey into multiple observations in order to maximize the probability of detecting a burst.
Since the rate scaling is dependent on distance,
choosing sources closer than FRB121102 is advised. 
Finally, 
SLSNe/LGRBs with coincident persistent radio sources, like PTF10hgi, should be
the primary sources to observe for future surveys of this kind; they should
preferably be observed at higher frequencies, as we cannot be certain that the
SNR is transparent at L-band.
The UWL might be the ideal instrument for following up on this work for two
reasons: 
\textit{i)} The SNR of the targets observed here are most likely transparent in at least
the upper part of UWL's band, making the optically thick-thin transition 
potentially observable with a single receiver.
\textit{ii)} The scaled FRB121102 UWL rates are higher than the ones for the S45mm receiver.
This implies that the time needed to constrain the 95\% CL upper rate limits of the targets observed
in this work with the UWL is much less than for the S45mm receiver. 
We show in Table \ref{tab:UWL} that these times range between 1--16 hrs. 

Recent localisations of FRBs 
\citep{Bannistereaaw5903,2019Natur.572..352R,2019Sci...366..231P,2020Natur.577..190M} 
have revealed host
galaxies differing from FRB121102, with them being lenticular or spiral in shape,
and more massive. 
The localisation of FRB180916.J1058+65 \citep{2020Natur.577..190M}
is of particular interest, as it is the only other localised repeating FRB.
The host of FRB180916.J1058+65 is a spiral galaxy and is both more massive and 
has higher metallicity than the host of FRB121102, 
rendering it different to hosts of SLSNe/LGRBs as well.
This bursting source also has no persistent radio counterpart, and
the burst absolute RM value is roughly $100$~rad~m$^{-2}$, three orders of magnitude lower
than FRB121102. The two bursting sources are however both localised within
star forming regions of their respective host galaxies.
FRB180916.J1058+65 still fits within the framework of a magnetar embedded
in an SNR if the system is a few hundred years old \citep{2020Natur.577..190M}. By then the 
persistent radio source would have faded and the RM decreased to the
observed value.
This begs the question whether or not the host galaxy of FRB121102
is a typical host of repeating FRBs.
Expanding future surveys like in this work to include galaxies similar to
hosts of other localised FRBs could be more fruitful.

\begin{table}
    \begin{center}
    \caption{Burst rates of an FRB121102-like source located at the targets 
            observed in this work.
            \textit{From left to right}:
            Source name, burst rate of an FRB121102-like source scaled to the 
            UWL receiver, and the 
            observing time required to constrain those rates at the 95\% confidence level.}
    \label{tab:UWL}
    \begin{tabular}{c c c}
    Source name & $R_\text{FRB121102}$ [hr$^{-1}$] & $T_\text{constr}$ [hr] \\
    \hline \\ [-1.0em]
    GRB050826   &0.3    &11 \\
    GRB051109B  &3.2    &1 \\
    GRB111225A  &0.3    &11 \\ 
    PTF09cnd    &0.4    &8 \\
    PTF10uhf    &0.3    &10 \\
    PTF10bjp    &0.2    &16 \\
    SN2010gx    &0.4    &7 \\
    PTF12dam    &1.9    &2 \\
    LSQ12dlf    &0.4    &8 \\
    PTF10hgi    &2.2    &1 
    \end{tabular}
    \end{center}
\end{table}

\section*{Acknowledgements}
% Entry for the table of contents, for this guide only
%\addcontentsline{toc}{section}{Acknowledgements}
Based on observations with the 100-m telescope of the MPIfR (Max-Planck-Institut f{\"u}r Radioastronomie) at Effelsberg.
Some of the results of this paper have been derived using the
\texttt{FRUITBAT} package \citep{2019JOSS....4.1399B}.
The Parkes radio telescope is funded by the Commonwealth of Australia
for operation as a National Facility managed by CSIRO. The time for
this project was allocated from the Director's Discretionary Time under
the Project ID PX048.
We thank Vincent Morello for assisting with our Parkes observations,
and C.~R.~H.~Walker for excellent comments and discussions.
LGS is a Lise Meitner independent research group leader and acknowledges support from the Max Planck Society.
We thank the referees for their comments which helped 
improving this manuscript.
%thanks

%%%%%%%%%%%%%%%%%%%%%%%%%%%%%%%%%%%%%%%%%%%%%%%%%%

%%%%%%%%%%%%%%%%%%%% REFERENCES %%%%%%%%%%%%%%%%%%

% The best way to enter references is to use BibTeX:

\bibliographystyle{mnras}
\bibliography{bibby}

%%%%%%%%%%%%%%%%%%%%%%%%%%%%%%%%%%%%%%%%%%%%%%%%%%

%%%%%%%%%%%%%%%%% APPENDICES %%%%%%%%%%%%%%%%%%%%%

%\appendix
%\section{Journal abbreviations}
%\label{sec:abbreviations}

%\clearpage % to avoid the long table breaking up the formatting examples
%\section{Advanced formatting examples}
%\label{sec:advanced}

%%%%%%%%%%%%%%%%%%%%%%%%%%%%%%%%%%%%%%%%%%%%%%%%%%

% Don't change these lines
\bsp	% typesetting comment
\label{lastpage}
\end{document}